\documentstyle[sprocl]{article}

\newcommand{\VEV}[1]{\left\langle{#1}\right\rangle}

\newcommand{\ket}[1]{\,\left|\,{#1}\right\rangle}
\newcommand{\longvec}[1]{\overrightarrow{\!\!#1}}
\renewcommand{\bar}[1]{\overline{#1}}
\newcommand{\etal} {{\em et al.}}
\newcommand{\ie}   {{\em i.e.}}
\newcommand{\eg}   {{\em e.g.}}
\newcommand{\M}    {{\cal M}}
\newcommand{\R}    {{\cal R}}
\input{epsf}

\begin{document}

\title{QCD PHENOMENA AND THE LIGHT-CONE WAVEFUNCTIONS OF HADRONS}

\author{S. J. BRODSKY}

\address{Stanford Linear Accelerator Center\\
Stanford University, Stanford, CA 94309\\
e-mail: sjbth@slac.stanford.edu}

\maketitle
\abstracts{
Light-cone Fock-state wavefunctions encode the properties of a hadron in terms
of its fundamental quark and gluon degrees of freedom. A recent experiment at
Fermilab, E791, demonstrates that the color coherence and shape of
light-cone wavefunctions in longitudinal momentum fraction can be directly
measured by the high energy diffractive jet dissociation of hadrons on nuclei.
Given the proton's light-cone wavefunctions, one can compute not
only the quark and gluon distributions measured in deep inelastic lepton-proton
scattering, but also the multi-parton correlations
which control the distribution of particles in the proton fragmentation
region and dynamical higher twist effects.  First-principle predictions can be
made for structure functions at small and large light-cone momentum
fraction $x$. Light-cone wavefunctions also provide a systematic framework for
evaluating exclusive hadronic matrix elements, including timelike heavy hadron decay
amplitudes and form factors. In principle, light-cone wavefunctions can be
computed in nonperturbative QCD by diagonalizing the light-cone Hamiltonian
using the DLCQ method, as in dimen\-sionally reduced collinear QCD.}

\section{Introduction}

In a relativistic collision, the incident hadron projectile
presents itself as an  ensemble of coherent states containing various numbers
of quark and gluon quanta. Thus when a laser beam crosses a
proton at fixed ``light-cone" time
$\tau = t+z/c= x^0 + x^z$, it
encounters a baryonic state with a given number of quarks, anti-quarks, and
gluons in flight  with  $n_q - n_{\bar q} = 3$.  The natural formalism for
describing these hadronic components in QCD is the light-cone Fock
representation obtained by quantizing the theory at fixed
$\tau$.\cite{PinskyPauli}
 For example, the proton state has the Fock expansion
\begin{eqnarray}
\ket p &=& \sum_n \VEV{n\,|\,p}\, \ket n \nonumber \\
&=& \psi^{(\Lambda)}_{3q/p} (x_i,\vec k_{\perp i},\lambda_i)\,
\ket{uud} \\[1ex]
&&+ \psi^{(\Lambda)}_{3qg/p}(x_i,\vec k_{\perp i},\lambda_i)\,
\ket{uudg} + \cdots \nonumber
\label{eq:b}
\end{eqnarray}
representing the expansion of the exact QCD eigenstate on a non-interacting
quark and gluon basis.  The probability amplitude
for each such
$n$-particle state of on-mass shell quarks and gluons in a hadron is given by a
light-cone Fock state wavefunction
$\psi_{n/H}(x_i,\vec k_{\perp i},\lambda_i)$, where the constituents have
longitudinal light-cone momentum fractions
\begin{equation}
x_i = \frac{k^+_i}{p^+} = \frac{k^0+k^z_i}{p^0+p^z}\ ,\quad \sum^n_{i=1} x_i= 1
\ ,
\label{eq:c}
\end{equation}
relative transverse momentum
\begin{equation}
\vec k_{\perp i} \ , \quad \sum^n_{i=1}\vec k_{\perp i} = \vec 0_\perp \ ,
\label{eq:d}
\end{equation}
and helicities $\lambda_i.$  The effective lifetime of each configuration
in the laboratory frame is ${2 P_{lab}\over{\M}_n^2- M_p^2} $ where
\begin{equation} 
\M^2_n = \sum^n_{i=1} \frac{k^2_\perp + m^2}{x} < \Lambda^2    \label{eq:a}
\end{equation} is the off-shell invariant mass and $\Lambda$ is a global
ultraviolet regulator.  The form of the $\psi^{(\Lambda)}_{n/H}(x_i,
\vec k_{\perp i},\Lambda_c)$ is invariant under longitudinal boosts; \ie,\ the
light-cone wavefunctions expressed in the relative coordinates $x_i$ and
$k_{\perp i}$   are independent of the total momentum
$P^+$,
$\vec P_\perp$ of the hadron.

Thus the interactions of the proton reflects an average over the interactions 
of its fluctuating states. For example, a valence state with small impact separation, and
thus a small color dipole moment, would be expected  to interact weakly in a
hadronic or nuclear target reflecting its color transparency.  The nucleus
thus filters differentially different hadron
components.\cite{Bertsch,MillerFrankfurtStrikman}  The ensemble
\{$\psi_{n/H}$\} of such light-cone Fock
wavefunctions is a key concept for hadronic physics, providing a conceptual
basis for representing physical hadrons (and also nuclei) in terms of their
fundamental quark and gluon degrees of freedom.  Given the
$\psi^{(\Lambda)}_{n/H},$ we can construct any spacelike electromagnetic or
electroweak form factor from the diagonal overlap of the LC wavefunctions.\cite{BD}
Similarly, the matrix elements of the currents that define quark and gluon
structure functions can be computed from the integrated squares of the LC
wavefunctions.\cite{LB,BrodskyLepage}

It is thus important to not only compute the
spectrum of hadrons and gluonic states, but also to
determine the wavefunction of each QCD bound state in terms of its
fundamental quark and gluon degrees of freedom. If we could obtain such
nonperturbative solutions of QCD, then we could compute the quark and
gluon structure functions and distribution amplitudes which control
hard-scattering inclusive and exclusive reactions as well as calculate the
matrix elements of currents which underlie electroweak form factors and the weak
decay amplitudes of the light and heavy hadrons.  The light-cone wavefunctions
also determine the multi-parton correlations  which control the distribution of
particles in the proton fragmentation region as well as dynamical higher twist
effects.  Thus one can analyze not only the deep inelastic structure
functions but also the fragmentation of the spectator system. 
Knowledge of hadron wavefunctions would also open a window to a deeper
understanding of the physics of QCD at the amplitude level, illuminating exotic
effects of the theory such as color transparency, intrinsic heavy quark
effects, hidden color, diffractive processes, and the QCD van der Waals interactions.

Solving a quantum field theory such as QCD is clearly not easy.  However, highly
nontrivial, one-space one-time relativistic quantum field theories which
mimic many of the features of QCD, have already been completely solved
using light-cone
Hamiltonian methods.\cite{PinskyPauli}  Virtually any
(1+1) quantum field theory can be solved using the method of Discretized
Light-Cone-Quantization (DLCQ). \cite{DLCQ,Schlad}  In DLCQ, the Hamiltonian
$H_{LC}$, which can be constructed from the Lagrangian using light-cone time
quantization, is completely diagonalized, in analogy to Heisenberg's solution of
the eigenvalue problem in quantum mechanics. The quantum field theory problem is
rendered discrete by imposing periodic or anti-periodic boundary
conditions.  The eigenvalues and eigensolutions of collinear QCD then give the
complete spectrum of hadrons, nuclei, and gluonium and their respective
light-cone
wavefunctions.  A beautiful
example is ``collinear" QCD:  a variant of $QCD(3+1)$ defined by
dropping all of interaction terms in $H^{QCD}_{LC}$ involving
transverse momenta.\cite{Kleb} Even though this theory is effectively
two-dimensional, the transversely-polarized degrees of freedom of the gluon
field are retained as two scalar fields.  Antonuccio and Dalley
\cite{AD} have used DLCQ to solve this theory. The diagonalization of
$H_{LC}$ provides not only the complete bound and continuum spectrum of
the collinear theory, including the gluonium states, but it also yields the
complete ensemble of light-cone Fock state wavefunctions needed to construct
quark and gluon structure functions for each bound state.  Although the
collinear theory is a drastic approximation to physical $QCD(3+1)$, the
phenomenology of its DLCQ solutions demonstrate general gauge
theory features, such as the peaking of the wavefunctions at minimal
invariant mass, color
coherence and the helicity retention of leading partons in the polarized
structure functions at $x\rightarrow 1$. 
The solutions of the quantum field theory can be obtained for arbitrary
coupling strength, flavors, and colors.

The light-cone Fock formalism is defined in the following way:  one
first constructs the light-cone time evolution operator $P^-=P^0-P^z$
and the invariant mass operator $H_{LC}= P^- P^+-P^2_\perp $ in
light-cone gauge $A^+=0$ from the QCD Lagrangian.
The total longitudinal momentum $P^+ = P^0 + P^z$ and transverse
momenta $\vec P_\perp$ are conserved, \ie\ are independent of the interactions.
The matrix elements of
$H_{LC}$ on the complete orthonormal basis $\{\ket{n}\}$ of the free
theory $H^0_{LC} =
H_{LC}(g=0)$ can then be constructed.  The matrix elements
$\VEV{n\,|\,H_{LC}\,|\,m}$ connect Fock states differing by 0,
1, or 2 quark or gluon quanta, and they include the instantaneous quark
and gluon contributions imposed by eliminating dependent degrees of
freedom in light-cone gauge.
In the discretized light-cone method (DLCQ), the matrix
elements\break
$\VEV{n\,|\,H^{\Lambda)}_{LC}\,|\,m}$, are made discrete in momentum space by
imposing periodic or anti-periodic boundary conditions in $x^-=x^0 - x^z$ and
$\vec x_\perp$. Upon diagonalization of $H_{LC}$, the eigenvalues provide the
invariant mass of the bound states and eigenstates of the continuum.

In practice it is essential to introduce an
ultraviolet regulator in order to limit the total range of
$\VEV{n\,|\,H_{LC}\,|\,m}$, such as the ``global" cutoff in the invariant
mass of the free Fock state.
One can also introduce a ``local" cutoff to limit
the change in invariant mass $|\M^2_n-\M^2_m| < \Lambda^2_{\rm local}$
which provides spectator-independent regularization of the
sub-divergences associated with mass and coupling renormalization.
Recently, Hiller,  McCartor,  and I have shown\cite{BHM} that the Pauli-Villars
method has advantages for regulating light-cone quantized Hamitonian
theory. We show that  Pauli-Villars fields satisfying three spectral
conditions will regulate the interactions in the ultraviolet, while at same time
avoiding spectator-dependent renormalization and preserving chiral
symmetry.

The natural renormalization scheme for the QCD coupling is $\alpha_V(Q)$, the
effective charge defined from the scattering of two infinitely-heavy
quark test charges.  The renormalization scale can then be determined
from the virtuality of the exchanged momentum, as in the BLM and
commensurate scale methods.\cite{BLM,CSR,BrodskyKataevGabaladzeLu,BJPR}

In principle, we could also construct the wavefunctions of QCD(3+1)
starting with collinear QCD(1+1) solutions by systematic perturbation
theory in $\Delta H$, where $\Delta H$ contains the terms which
produce particles at non-zero $k_\perp$, including the terms linear and
quadratic in the transverse momenta $\longvec k_{\perp i}$ which are
neglected in the Hamilton $H_0$ of collinear QCD.  We can write the exact
eigensolution of the full Hamiltonian as
\[ \psi_{(3+1)} = \psi_{(1+1)} + \frac{1}{M^2-H + i \epsilon }\,
\Delta H\, \psi_{(1+1)} \ , \]
where
\[\frac{1}{M^2-H + i \epsilon }
 = {\frac{1}{M^2-H_0 + i \epsilon }} +
{\frac{1}{M^2-H+ i \epsilon }}\Delta H{\frac{1}{M^2-H_0 + i \epsilon }} \]
can be represented as the continued iteration of the Lippmann Schwinger
resolvant.
Note that the matrix
$(M^2-H_0)^{-1}$ is known to any desired precision from the DLCQ solution
of collinear QCD.

\section{Electroweak Matrix Elements}

\vspace{.5cm}
\begin{figure}[htb]
\begin{center}
\leavevmode
\epsfbox{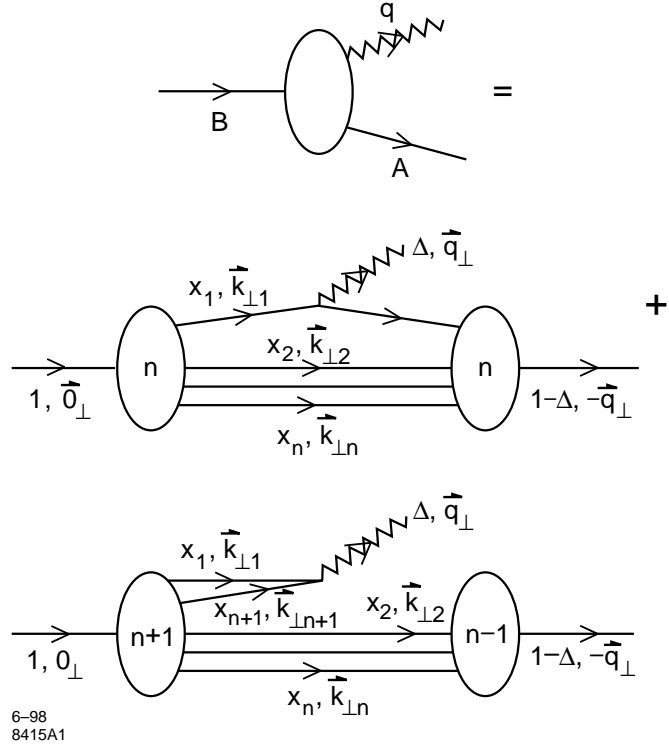}
\end{center}
\caption[*]{Exact representation of electroweak decays and time-like form factors in the
light-cone Fock representation.
}
\label{fig1}
\end{figure}

Dae Sung Hwang and I have recently shown that
exclusive semileptonic $B$-decay amplitudes, such as
$B\rightarrow A \ell \bar{\nu}$ can be evaluated exactly in the light-cone formalism.
\cite{BrodskyHwang} These timelike decay matrix elements require the computation of
the diagonal matrix element $n \rightarrow n$ where parton number is conserved, 
and the off-diagonal $n+1\rightarrow n-1$ convolution where the current operator 
annihilates a $q{\bar{q'}}$ pair in the initial $B$
wavefunction. See Fig. \ref{fig1}.  This term is a consequence of the fact that the
time-like decay $q^2 = (p_\ell + p_{\bar{\nu}} )^2 > 0$
requires a positive light-cone momentum fraction
$q^+ > 0$.  Conversely for space-like currents, one can choose $q^+=0$, as in
the Drell-Yan-West representation of the space-like electromagnetic form
factors. However, as can be  seen from the explicit analysis of the form factor in a
perturbation model, the off-diagonal convolution can yield a nonzero $q^+/q^+$
limiting form as $q^+ \rightarrow 0$.  This extra term appears specifically in the case of ``bad"
currents such as $J^-$ in which the coupling to $q\bar q$ fluctuations in the light-cone
wavefunctions are favored.  In effect, the $q^+ \rightarrow 0$ limit generates
$\delta(x)$ contributions as residues of the $n+1\rightarrow n-1$ contributions.  
The necessity for this zero mode $\delta(x)$ terms has been noted by
Chang, Root and Yan,\cite{CRY} and Burkardt.\cite{BUR}

The off-diagonal $n+1 \rightarrow n-1$ contributions provide a new
perspective for the physics of $B$-decays.  A semileptonic decay involves not only matrix
elements where a quark changes flavor, but also a contribution where the leptonic pair is
created from the annihilation of a $q {\bar{q'}}$ pair within the Fock states of the initial $B$
wavefunction.  The semileptonic decay thus can occur from the annihilation of a
nonvalence quark-antiquark pair in the initial hadron.  This feature will carry
over to exclusive hadronic $B$-decays, such as $B^0 \rightarrow
\pi^-D^+$.  In this case the pion can be produced from the coalescence of a
$d\bar u$ pair emerging from the initial higher particle number Fock
wavefunction of the $B$.  The $D$ meson is then formed from the remaining quarks
after the internal exchange of a $W$ boson.

In principle, a precise evaluation of the hadronic matrix elements needed for $B$-decays
and other exclusive electroweak decay amplitudes requires knowledge of all of
the light-cone Fock wavefunctions of the initial and final state hadrons.
In the case of model gauge theories such as QCD(1+1)\cite{Horn}  or collinear
QCD \cite{AD} in one-space and one-time dimensions, the complete evaluation of
the light-cone wavefunction is possible for each baryon or meson bound-state
using the DLCQ  method.  It would be interesting to use such solutions as a model 
for physical $B$-decays.

The existence of an exact formalism
provides a basis for systematic approximations and a control over neglected
terms.  For example, one can analyze exclusive semileptonic
$B$-decays  which  involve hard internal momentum transfer using a
perturbative QCD formalism patterned after the analysis of form factors at
large momentum transfer.\cite{LB}    The hard-scattering analysis proceeds by
writing each hadronic wavefunction as a sum of soft and hard contributions
\begin{equation}
\psi_n = \psi^{{\rm soft}}_n (\M^2_n < \Lambda^2) + \psi^{{\rm hard}}_n
(\M^2_n >\Lambda^2) ,
\end{equation}
where $\M^2_n $  is the invariant mass of the partons in the $n$-particle Fock state and
$\Lambda$ is the separation scale.
The high internal momentum contributions to the wavefunction $\psi^{{\rm
hard}}_n $ can be calculated systematically from QCD perturbation theory
by iterating  the gluon exchange kernel.  The contributions from  high
momentum transfer exchange to the
$B$-decay amplitude can then be written as a convolution of a hard scattering
quark-gluon scattering amplitude $T_H$  with the distribution
amplitudes $\phi(x_i,\Lambda)$, the valence wavefunctions obtained by integrating the
constituent momenta  up to the separation scale
${\cal M}_n < \Lambda < Q$.  This is the basis for the perturbative hard
scattering analyses.\cite{BHS,Sz,BALL,BABR} 
 In the exact analysis, one can
identify the hard PQCD contribution as well as the soft contribution from
the convolution of the light-cone wavefunctions. Furthermore, the hard
scattering contribution can be systematically improved.  For example, off-shell
effects can be retained in the evaluation of
$T_H$ by utilizing the exact light-cone energy denominators.

More generally, hard  exclusive hadronic amplitudes such as quarkonium
decay, heavy hadron decay,  and  scattering amplitudes where the hadrons
are scattered with  momentum transfer can be factorized  as the convolution of
the light-cone Fock state wavefunctions with quark-gluon matrix elements
\cite{LB}
\begin{eqnarray}
\M_{\rm Hadron} &=& \prod_H \sum_n \int
\prod^{n}_{i=1} d^2k_\perp \prod^{n}_{i=1}dx\, \delta
\left(1-\sum^n_{i=1}x_i\right)\, \delta
\left(\sum^n_{i=1} \vec k_{\perp i}\right) \nonumber \\[2ex]
&& \times \psi^{(\Lambda)}_{n/H} (x_i,\vec k_{\perp i},\Lambda_i)\,
T_H^{(\Lambda)} \ .
\label{eq:e}
\end{eqnarray}
Here $T_H^{(\Lambda)}$ is the underlying quark-gluon
subprocess scattering amplitude, where the (incident or final) hadrons are
replaced by quarks and gluons with momenta $x_ip^+$, $x_i\vec
p_{\perp}+\vec k_{\perp i}$ and invariant mass above the
separation scale $\M^2_n > \Lambda^2$.
The essential part of the wavefunction is the hadronic distribution amplitudes,
\cite{LB} defined as the integral over transverse momenta of the valence (lowest
particle number) Fock wavefunction; \eg\ for the pion
\begin{equation}
\phi_\pi (x_i,Q) \equiv \int d^2k_\perp\, \psi^{(Q)}_{q\bar q/\pi}
(x_i, \vec k_{\perp i},\lambda)
\label{eq:f}
\end{equation}
where the global cutoff $\Lambda$ is identified with the
resolution $Q$.  The distribution amplitude controls leading-twist exclusive
amplitudes at high momentum transfer, and it can be related to the
gauge-invariant Bethe-Salpeter wavefunction at equal light-cone time
$\tau = x^+$.  The $\log Q$ evolution of the hadron distribution amplitudes
$\phi_H (x_i,Q)$ can be derived from the
perturbatively-computable tail of the valence light-cone wavefunction in the
high transverse momentum regime.\cite{LB}   In general the LC ultraviolet
regulators provide a factorization scheme for elastic and inelastic
scattering, separating the hard dynamical contributions with invariant mass
squared $\M^2 > \Lambda^2_{\rm global}$ from the soft physics with
$\M^2 \le \Lambda^2_{\rm global}$ which is incorporated in the
nonperturbative LC wavefunctions.  The DGLAP evolution of quark and gluon
distributions can also be derived by computing the variation of the Fock
expansion with respect to $\Lambda^2$.\cite{LB}

Given the solution
for the hadronic wavefunctions $\psi^{(\Lambda)}_n$ with $\M^2_n <
\Lambda^2$, one can construct the wavefunction in the hard regime with
$\M^2_n > \Lambda^2$ using projection operator techniques.\cite{LB} The
construction can be done perturbatively in QCD since only high invariant mass,
far off-shell matrix elements are involved.  One can use this method to
derive the physical properties of the LC wavefunctions and their matrix elements
at high invariant mass.  Since $\M^2_n = \sum^n_{i=1}
\left(\frac{k^2_\perp+m^2}{x}\right)_i $, this method also allows the derivation
of the asymptotic behavior of light-cone wavefunctions at large $k_\perp$, which
in turn leads to predictions for the fall-off of form factors and other
exclusive
matrix elements at large momentum transfer, such as the quark counting rules
for predicting the nominal power-law fall-off of two-body scattering amplitudes
at fixed
$\theta_{cm}.$\cite{BrodskyLepage} The phenomenological successes of these rules
can be understood within QCD if the coupling
$\alpha_V(Q)$ freezes in a range of relatively
small momentum transfer.\cite{BJPR}

\section{Measurement of Light-cone  Wavefunctions via Diffractive
Dissociation.} Diffractive multi-jet production in heavy
nuclei provides a novel way to measure the shape of the LC Fock
state wavefunctions and test color transparency. For example, consider the reaction
\cite{Bertsch,MillerFrankfurtStrikman}
$\pi A \rightarrow {\rm Jet}_1 + {\rm Jet}_2 + A^\prime$
at high energy where the nucleus $A^\prime$ is left intact in its ground
state.  The transverse momenta of the jets have to balance so that
$
\vec k_{\perp i} + \vec k_{\perp 2} = \vec q_\perp < \R^{-1}_A \ ,
$
and the light-cone longitudinal momentum fractions have to add to
$x_1+x_2 \sim 1$ so that $\Delta p_L < R^{-1}_A$.  The process can
then occur coherently in the nucleus.  Because of color transparency,  i.e.,
the cancelation of color interactions in a small-size color-singlet
hadron,  the valence wavefunction of the pion with small impact
separation will penetrate the nucleus with minimal interactions,
diffracting into jet pairs. \cite{Bertsch}
The $x_1=x$, $x_2=1-x$ dependence of
the di-jet distributions will thus reflect the shape of the pion distribution
amplitude; the $\vec k_{\perp 1}- \vec k_{\perp 2}$
relative transverse momenta of the jets also gives key information on
the underlying shape of the valence pion wavefunction.  The QCD analysis can be
confirmed by the observation that the diffractive nuclear amplitude
extrapolated to $t = 0$ is linear in nuclear number $A$, as predicted by
QCD color
transparency.  The integrated diffractive rate should scale as $A^2/R^2_A \sim
A^{4/3}$. A diffractive dissociation experiment of this type, E791,  is now in
progress at Fermilab using 500 GeV incident pions on nuclear
targets.\cite{E791}  The preliminary results from E791 appear to be consistent
with color transparency.  The momentum fraction distribution of the jets is
consistent with a valence light-cone wavefunction of the pion consistent with
the shape of the asymptotic distribution amplitude. Data from CLEO for the
$\gamma
\gamma^* \rightarrow \pi^0$ transition form factor also favor a form for
the pion distribution amplitude close to the asymptotic solution \cite{LB}
$\phi^{\rm asympt}_\pi (x) =
\sqrt 3 f_\pi x(1-x)$ to the perturbative QCD evolution
equation.\cite{Kroll,Rad,BJPR} 
It will also be interesting to study diffractive tri-jet production using proton
beams
$ p A \rightarrow {\rm Jet}_1 + {\rm Jet}_2 + {\rm Jet}_3 + A^\prime $ to
determine the fundamental shape of the 3-quark structure of the valence
light-cone wavefunction of the nucleon at small transverse separation.
Conversely, one can use incident real and virtual photons:
$ \gamma^* A \rightarrow {\rm Jet}_1 + {\rm Jet}_2 + A^\prime $ to
confirm the shape of the calculable light-cone wavefunction for
transversely-polarized and longitudinally-polarized virtual photons.  Such
experiments will open up a remarkable, direct window on the amplitude
structure of hadrons at short distances.

\section{Other Applications of Light-Cone Quantization to QCD Phenomenology}

{\it Diffractive vector meson photoproduction.} The
light-cone Fock wavefunction representation of hadronic amplitudes
allows a simple eikonal analysis of diffractive high energy processes, such as
$\gamma^*(Q^2) p \to \rho p$, in terms of the virtual photon and the vector
meson Fock state light-cone wavefunctions convoluted with the $g p \to g p$
near-forward matrix element.\cite{BGMFS} One can easily show that only small
transverse size $b_\perp \sim 1/Q$ of the vector meson distribution
amplitude  is involved. The hadronic interactions are minimal, and thus the
$\gamma^*(Q^2) N \to
\rho N$ reaction can occur coherently throughout a nuclear target in reactions
without absorption or shadowing. The $\gamma^* A \to V A$ process thus
provides a natural framework for testing QCD color transparency.\cite{BM}  Evidence for
color transparency in such reactions has been found by Fermilab experiment
E665.

{\it Regge behavior of structure functions.} The light-cone wavefunctions
$\psi_{n/H}$ of a hadron are not independent of each other, but rather are
coupled via the equations of motion. Antonuccio, Dalley and I
\cite{ABD} have used the constraint of finite ``mechanical'' kinetic energy to
derive ``ladder relations" which interrelate the light-cone wavefunctions of
states differing by one or two gluons.  We then use these relations to derive the
Regge behavior of both the polarized and unpolarized structure functions at $x
\rightarrow 0$, extending Mueller's derivation of the BFKL hard
QCD pomeron from the properties of heavy quarkonium light-cone wavefunctions at
large $N_C$ QCD.\cite{Mueller}

{\it Structure functions at large $x_{bj}$.} The behavior of structure functions
where one quark has the entire momentum requires the knowledge of LC
wavefunctions
with $x \rightarrow 1$ for the struck quark and $x \rightarrow 0$ for the
spectators.  This is a highly off-shell configuration, and thus one can
rigorously derive quark-counting and helicity-retention rules for the power-law behavior of
the polarized and unpolarized quark and gluon distributions in the $x
\rightarrow 1$ endpoint domain.  It is interesting to note that the evolution
of structure functions is minimal in this domain because the struck quark
is highly virtual as $x\rightarrow 1$; \ie\ the starting point $Q^2_0$ for evolution
cannot be held fixed, but must be larger than a scale of order
$(m^2 + k^2_\perp)/(1-x)$.\cite{LB,BrodskyLepage,Dmuller}

{\it Intrinsic gluon and heavy quarks.}
The main features of the heavy sea quark-pair contributions of the Fock
state expansion of light hadrons can also be derived from perturbative QCD,
since $\M^2_n$ grows with
$m^2_Q$.  One identifies two contributions to the heavy quark sea, the
``extrinsic'' contributions which correspond to ordinary gluon splitting, and
the ``intrinsic" sea which is multi-connected via gluons to the valence quarks.
The intrinsic sea is thus sensitive to the hadronic bound state
structure.\cite{IC}  The maximal contribution of the
intrinsic heavy quark occurs at $x_Q \simeq {m_{\perp Q}/ \sum_i m_\perp}$
where $m_\perp = \sqrt{m^2+k^2_\perp}$;
\ie\ at large $x_Q$, since this minimizes the invariant mass $\M^2_n$.  The
measurements of the charm structure function by the EMC experiment are
consistent with intrinsic charm at large $x$ in the nucleon with a
probability of order $0.6 \pm 0.3 \% $.\cite{HSV} Similarly, one can distinguish intrinsic
gluons which are associated with multi-quark interactions and extrinsic gluon
contributions associated with quark substructure.\cite{BS} One can also use this
framework to isolate the physics of the anomaly contribution to the Ellis-Jaffe
sum rule.

{\it Materialization of far-off-shell configurations.}
In a high energy hadronic collisions, the highly-virtual states of a hadron can be
materialized into physical hadrons simply by the soft interaction of any of the
constituents.\cite{BHMT}  Thus a proton state with intrinsic charm $\ket{ u
u d \bar  c c}$ can be materialized, producing a $J/\psi$ at large $x_F$,  by the
interaction of a light-quark in the target.  The production
occurs on the front-surface of a target nucleus, implying an $A^{2/3}$
$J/\psi$ production cross section at large $x_F$ , which is consistent with
experiment, such as Fermilab experiments E772 and  E866.

{\it Rearrangement mechanism in heavy quarkonium decay.}
It is usually assumed that a heavy quarkonium state such as the
$J/\psi$ always decays to light hadrons via the annihilation of its heavy quark
constituents to gluons. However, as Karliner and I \cite{BK} have recently
shown, the transition  $J/\psi \to \rho
\pi$ can also occur by the rearrangement of the $c \bar c$ from the $J/\psi$
into the $\ket{ q \bar q c \bar c}$ intrinsic charm Fock state of the $\rho$ or
$\pi$. On the other hand, the overlap rearrangement integral in the
decay $\psi^\prime \to \rho \pi$ will be suppressed since the intrinsic
charm Fock state radial wavefunction of the light hadrons will evidently
not have nodes in its radial wavefunction. This observation provides a natural
explanation of the long-standing puzzle why the $J/\psi$ decays prominently to
two-body pseudoscalar-vector final states, whereas the $\psi^\prime$
does not.

{\it Asymmetry of intrinsic heavy quark sea.} As Burkardt and Warr\cite{Warr} first noted,
the higher Fock state of the proton $\ket{u u d s \bar s}$ should
resemble  a $\ket{ K \Lambda}$ intermediate state, since this minimizes its
invariant mass $\M$.  In such a state, the
strange quark has a higher mean momentum fraction $x$ than the $\bar
s$. \cite{Warr,Signal,BMa} Similarly, the helicity intrinsic strange
quark in this configuration will be anti-aligned with the helicity of the nucleon.%
\cite{Warr,BMa} This $Q \leftrightarrow \bar Q$ asymmetry
is a striking feature of the intrinsic heavy-quark sea.

{\it Comover phenomena.}
Light-cone wavefunctions describe not only the partons that interact in a hard
subprocess but also the associated partons freed from the projectile.  The
projectile partons which are comoving (\ie, which have similar rapidity) with
final state quarks and gluons can interact strongly producing (a) leading
particle effects, such as those seen in open charm hadroproduction; (b)
suppression of quarkonium\cite{BrodskyMueller} in favor of open heavy hadron
production, as seen in the E772 experiment; (c) changes in color configurations
and selection rules in quarkonium hadroproduction, as has been emphasized by
Hoyer and Peigne.\cite{HoyerPeigne}   All of these effects violate the
usual ideas of factorization for inclusive reactions. Further, more than one parton from the
projectile can enter the hard subprocess, producing dynamical higher twist
contributions, as seen for example in
Drell-Yan experiments.\cite{BrodskyBerger,Brandenburg}

{\it Jet hadronization in light-cone QCD.}
One of the goals of nonperturbative analysis in QCD is to compute jet
hadronization from first principles.  The DLCQ solutions provide a possible
method to accomplish this.  By inverting the DLCQ solutions, we can write the
``bare'' quark state of the free theory as
$\ket{q_0} = \sum \ket n \VEV{n\,|\,q_0}$
 where now $\{\ket n\}$ are the exact DLCQ eigenstates of
$H_{LC}$, and
$\VEV{n\,|\,q_0}$ are the DLCQ projections of the eigensolutions.  The expansion
in automatically infrared and ultraviolet regulated if we impose global cutoffs
on the DLCQ basis:
$\lambda^2 < \Delta\M^2_n < \Lambda^2 
$
where $\Delta\M^2_n = \M^2_n-(\Sigma \M_i)^2$.  It would be
interesting to study  jet hadronization at the amplitude level for
the existing DLCQ solutions to QCD (1+1) and collinear QCD.

\section*{Acknowledgments}
I wish to thank Misha Voloshin and his colleagues at the Theoretical
Physics Institute at the University of Minnesota for inviting me to
participate in this workshop which is dedicated to the memory of an
outstanding physicist and friend, Vlodya Gribov.   The analysis of timelike
decays presented in Section 2 is based on a collaboration with Dae Sung
Hwang.

\end{document}